\documentclass[]{elsarticle}

\usepackage[utf8]{inputenc}
\usepackage{amsmath}
\usepackage[linesnumbered]{algorithm2e}
\usepackage{amssymb}
\usepackage{tikz}

\newtheorem{definition}{Definition}
\newtheorem{lemma}{Lemma}

\newtheorem{example}{Example}

\begin{document}
	\begin{frontmatter}
		
		\title{PROMETHEE is Not Quadratic: \\An $\mathcal{O}(qn\log(n))$ Algorithm}
		\author[ua,ulb_code]{Toon Calders}
		\ead{toon.calders@uantwerpen.be}
		
		\author[ulb_code]{Dimitri Van Assche\corref{cor_p}}
		\ead{dvassche@ulb.ac.be}
		
        \address[ua]{Department of Mathematics and Computer Science, Universiteit Antwerpen}
		\address[ulb_code]{Computer \& Decision Engineering Department (CoDE), Université libre de Bruxelles}
		\cortext[cor_p]{Corresponding author}

		\begin{abstract}
        It is generally believed that the preference ranking method PROMETHEE has a quadratic time complexity. In this paper, however, we present an exact algorithm that computes PROMETHEE's net flow scores in time $\mathcal{O}(qn\log(n))$, where $q$ represents the number of criteria and $n$ the number of alternatives. The method is based on first sorting the alternatives after which the unicriterion flow scores of all alternatives can be computed in one scan over the sorted list of alternatives while maintaining a sliding window. This method works with the linear and level criterion preference functions. The algorithm we present is exact and, due to the sub-quadratic time complexity, vastly extends the applicability of the PROMETHEE method. Experiments show that with the new algorithm, PROMETHEE can scale up to millions of tuples.
		\end{abstract}
		
		
	\end{frontmatter}
	
	\section{Introduction}
	Multi-criteria decision aid (MCDA), that is, the study of simultaneously evaluating possible decisions on multiple conflicting criteria, has been an active research field for over 40 years. A common example in this area is that of buying a new car. When selecting which car to buy, one typically tries to minimize cost and consumption while maximizing performance, comfort, etc. Obviously no real car can be best on all those criteria at the same time. Therefore, the notion of optimal solution necessarily is replaced by that of a compromise solution~\cite{roy1981multicriteria}.
	
	Different methods to select a compromise solution from a set of alternatives have been proposed in the literature, which can be divided into 3 main categories \cite{roy2005paradigms}: those based upon multi-attribute utility theory (MAUT) \cite{figueira2005multiple}, outranking methods \cite{siskos1984outranking}, and interactive methods \cite{korhonen2005}.
	
	We only discuss the first two methods, as in the paper we only consider fully automatic ranking methods.
	In the methods based upon multi-attribute utility theory, scores are computed by aggregating, per alternative, its unicriterion utilities with functions that map a value into a preference degree. Most MAUT methods can therefore be computed efficiently, because the utility function is applied to each alternative for each criteria only once.
	
	On the other hand, outranking methods such as ELECTRE \cite{figueira2005electre} and PROMETHEE\footnote{Preference Ranking Organization METHod for Enrichment of Evaluations} \cite{brans2005} are based on pairwise comparisons of alternatives. These methods have a wide range of applications \cite{behzadian2010promethee}, yet are often criticized for their high complexity due to pairwise comparisons. Indeed, the straightforward computation of these ranking methods leads to a quadratic complexity in the number of alternatives. As a result, for a small number of alternative the straightforward implementation of these methods works fine, yet the performance degrades rapidly for an increasing number of alternatives.

In \cite{marinoni2006discussion}, a discussion is made about the performances of outranking methods in the context of geographical information analysis. Outranking methods are considered impractical when a large number of alternatives are involved, but there is no consensus on the meaning of large. Citing some concrete references, according to \cite{pereira1993multiple}: ``\textit{Outranking techniques [...] require pairwise or global comparisons among alternatives, which is obviously impractical for applications where the number of alternatives/cells in a database range in the tens or hundreds of thousands.}'' Furthermore, according to \cite{joerin2001using}, the problem already arises with more than 100 alternatives: ``\textit{[...] outranking methods have difficulties dealing with more than a hundred alternatives.}''
	In \cite{aerts2002using} it is grossly overgeneralized that the evaluation of a large number of alternatives is an issue of multi-criteria analysis: ``\textit{One major drawback of MCA is that it does not allow the comparison of a large number of alternatives. With only a few alternatives to be evaluated, it is almost certain that the best alternative chosen from the set is in fact a sub-optimal solution.}''
	
	In order to tackle these complexity problems, recently approximation methods have been developed to reduce the complexity of the calculation of PROMETHEE II by, for instance, the use of piecewise linear functions to approximate net flow scores of alternatives \cite{eppe2014approximating}. In this paper we show, however, that \textbf{efficient exact methods exist to compute PROMETHEE flow scores}. We present an exact method to compute flow scores of all alternatives considered for both PROMETHEE I and II, with time complexity $\mathcal{O}(qnlog(n))$ with $q$ the number of criteria and $n$ the number of alternatives. This is clearly a tremendous improvement over the $\mathcal{O}(qn^2)$ complexity of the straightforward computation by iterating over all pairs. Our method works for the two most popular preference functions: the linear and level criterion preference functions, and is based on sorting the alternatives for each criterion separately, followed by a linear scan over the sorted alternatives while maintaining a sliding window.
	
The paper is organized as follows. In Section \ref{sec:promethee}, we revisit the PROMETHEE ranking method. In Section \ref{sec:sb_algorithm}, we introduce our efficient sorting-based approach for calculating the flow scores of PROMETHEE, and establish its time complexity.  Finally, in Section~\ref{sec:results}, we compare the performance of the straightforward quadratic implementation with our new method, clearly illustrating the dramatic performance improvement, scaling up PROMETHEE to applications with millions of alternatives.
	
\section{PROMETHEE}
\label{sec:promethee}
In this section, we revisit the definition of the PROMETHEE I and II scoring methods. We refer the interested reader to \cite{brans2005} for a detailed description of PROMETHEE.
	
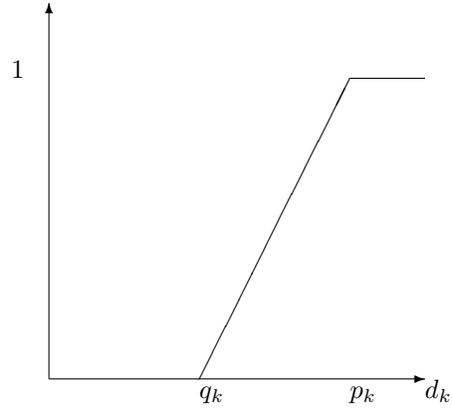
\begin{figure}
		\centering
		\setlength{\unitlength}{5cm}
		\begin{picture}(1,1)
		\put(0,0){\vector(0,1){1}}
		\put(0,0){\vector(1,0){1}}
		\put(0.8,0.8){\line(1,0){0.2}}
		\put(0.4,0){\line(1,2){0.4}}
		\put(-0.1,0.8){$1$}
		\put(1,-0.05){$d_k$}
		\put(0.8,-0.05){$p_k$}
		\put(0.4,-0.05){$q_k$}
		\end{picture}
		\caption[Linear preference function]{\label{lc} Linear preference function}
	\end{figure}

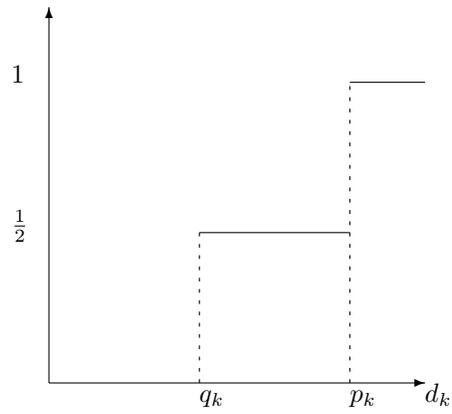
\begin{figure}
		\centering
		\setlength{\unitlength}{5cm}
		\begin{picture}(1,1)
		\put(0,0){\vector(0,1){1}}
		\put(0,0){\vector(1,0){1}}
		\put(0.8,0.8){\line(1,0){0.2}}
		\put(0.4,0.4){\line(1,0){0.4}}
		\multiput(0.4,0)(0, 0.03){14}{\line(0,1){0.01}}
		\multiput(0.8,0)(0, 0.03){27}{\line(0,1){0.01}}
		\put(-0.1,0.8){$1$}
		\put(-0.1,0.4){$\frac{1}{2}$}
		\put(1,-0.05){$d_k$}
		\put(0.8,-0.05){$p_k$}
		\put(0.4,-0.05){$q_k$}
		\end{picture}
		\caption[Level criterion preference function]{\label{lvlc} Level criterion preference function}
	\end{figure}

Let $A=\{a_1, a_2... , a_n\}$ be a set of $n$ alternatives and let $F = \{f_1, f_2... , f_q\}$ be a set of $q$ criteria. The evaluation of alternative $a_i$ for criterion $k$ will be denoted by a real value $f_k(a_i)$. We assume without loss of generality that a higher value for a criterion is better. For each pair of alternatives, we define $d_k (a_i, a_j)$ as the difference between $a_i$ and $a_j$ on criterion $k$.
	\begin{equation}
	d_k (a_i, a_j) := f_k(a_i) - f_k(a_j)
	\end{equation}
A preference function, denoted $P_k$, is associated with each criterion $k$. This function transforms the difference $d_k (a_i, a_j)$ between alternatives into a preference degree of $a_i$ over $a_j$ on criterion $k$. Multiple preference functions exist~\cite{brans2005}, such as the \emph{linear preference function} (Equation \ref{equ:pref_lin}, Figure \ref{lc}) and the level criterion preference function (Figure \ref{lvlc}).
In this paper we will consider the \emph{linear preference function}, which is defined as follows, given an indifference threshold $q_k$ and a preference threshold $p_k$:
	\begin{equation}
	\label{equ:pref_lin}
	P_k[d_k (a_i, a_j)] =
	\begin{cases}
	0	&	\text{if } d_k (a_i, a_j) \leq q_k \\
	\frac{d_k (a_i, a_j) - q_k}{p_k - q_k}	&	\text{if } q_k < d_k (a_i, a_j) \leq p_k \\
	1	& 	\text{if } p_k < d_k (a_i, a_j)
	\end{cases}
	\end{equation}	
The results in this paper can easily be extended to the level criterion preference function as well.
These two preference functions represent the most popular choices in the literature.

By applying the preference function $P_k$ to the difference $d_k$, we get the unicriterion preference degree $\pi_k$:
	\begin{equation}
	\pi_{k}(a_i,a_j):=P_{k}[d_{k}(a_i,a_j)]
	\end{equation}
The aggregated preference degree of alternative $a_i$ over $a_j$ is then computed as a weighted sum over the unicriterion preferences using weights $w_k$ associated with each criterion $k$. Weights are assumed to be positive and normalized.
\begin{equation}
\pi(a_i, a_j) := \sum_{k=1}^{q} w_k \pi_k(a_i, a_j)
\end{equation}
	
The last step consists in calculating the positive flow score denoted $\phi^{+} (a_i)$ and the negative flow score denoted $\phi^{-} (a_i)$ which are combined into the final flow score of $a_i$ as follows:
	\begin{eqnarray}
  \phi^{+} (a_i) &:=& \frac{1}{n-1} \sum_{x \in A} \pi(a_i, x)
\\\phi^{-} (a_i) &:=& \frac{1}{n-1} \sum_{x \in A} \pi(x, a_i)
\\\phi (a_i)     &:=& \phi_{A}^{+} (a_i) - \phi_{A}^{-} (a_i)
	\end{eqnarray}
	
The PROMETHEE I (partial) ranking is obtained as the intersection of the rankings induced by $\phi^{+}$ and $\phi^{-}$. PROMETHEE II gives a complete ranking induced by $\phi$. For an interpretation of the net flow scores, the interested reader is referred to \cite{mareschal2008rank}.
	
In this paper we will use the following, equivalent definition of the flow score:	
	\begin{eqnarray}
		\phi_k^+(a_i)&:=&\frac{1}{n-1}\sum_{x\in A} \pi_k(a_i,x)
		\\\phi_k^-(a_i)&:=&\frac{1}{n-1}\sum_{x\in A} \pi_k(x,a_i)
        \\\phi_k(a_i)&:=&\phi_k^+(a)-\phi_k^-(a_i)
        \\\phi(a_i)&=&\sum_{k=1}^q w_k \phi_k(a_i)
	\end{eqnarray}
	
	$\phi^{+}_{k} (a_i)$ and $\phi^{-}_{k} (a_i)$ are respectively called the \emph{unicriterion positive flow score} and \emph{unicriterion negative flow score} of the alternative $a_i$ on criterion $k$.
	

\section{Sorting-Based Algorithm for PROMETHEE}
	\label{sec:sb_algorithm}
Our sorting-based algorithm is based on computing the unicriterion positive and negative flow scores $\phi_k^+(a)$ and $\phi_k^-(a)$ for all alternatives $a$ for each criterion $f_k$ separately. The time required for this computation will be shown to be $\mathcal{O}(n\log(n))$ per criterion. Hence, computing all unicriterion positive and negative flow scores for all criteria takes time $\mathcal{O}(qn\log(n))$. These scores can be combined into the final PROMETHEE II flow score in time $\mathcal{O}(qn)$, leading to an overall time complexity of $\mathcal{O}(qn\log(n))$.

	Given the symmetry of the expressions for $\phi_k^+(a)$ and $\phi_k^-(a)$, we will only elaborate on the computation of $\phi_k^+(a)$. Indeed, if we substitute $f_k(a)$ by $f'_k(a)=-f_k(a)$, the resulting ${\phi'}_k^+(a) = \phi^-_k(a)$. As such, exactly the same algorithm can be used to compute $\phi_k^+(a)$ for all $a$ as for computing $\phi_k^-(a)$ for all $a$.	In the following we only present the algorithm for the linear preference function, as the algorithm for the level criterion preference function is very similar.

To illustrate the sorting-based computation of the unicriterion positive flow function, we use the following running example.
	\begin{example}
		\label{ex:prelim}
		Consider the following table listing a number of destinations (the alternatives), together with their characteristics (the criteria) on the basis of which we want to decide where to go on vacation. We consider three criteria: hours of sunshine, price, and historical value, each of which have been marked on a numerical scale of 1 to 10.
		\begin{center}
			\begin{tabular}{|ll||c|c|c|}
				\hline
				&&$f_1$&$f_2$&$f_3$
                \\&Destination&Sunshine&Price&History
				\\\hline
				$a_1$ & Brussels  & 5     & 6     & 9
				\\$a_2$ & Paris     & 6     & 4     & 10
				\\$a_3$ & Blois     & 7     & 9     & 8
				\\$a_4$ & Berlin    & 7     & 8     & 8
				\\$a_5$ & Barcelona & 10    & 7     & 7
				\\\hline
			\end{tabular}
		\end{center}
		Furthermore for all $k=1,...,3$, the thresholds $q_k$ and $p_k$ are respectively $1$ and $3$. This means that a difference of $1$ score point or less will be ignored, and a difference of more than $3$ points will no longer give additional benefit. In between these two extremes, the contribution to the flow score will be linear.

With these parameters, the positive unicriterion flow score for criterion $f_1$ for alternative $a_4$ is:
\begin{multline*}
\phi^+_1(a_4)~=\frac{1}{4}(\pi_1(a_4,a_1)+\pi_1(a_4,a_2)+\pi_1(a_4,a_3)+\pi_1(a_4,a_4)+\pi_1(a_4,a_5))
\\=~\frac{1}{4}\left(\frac{1}{2}+0+0+0+0\right)~=~\frac{1}{8}
\end{multline*}
	\end{example}
	
\subsection{Window of an Alternative}
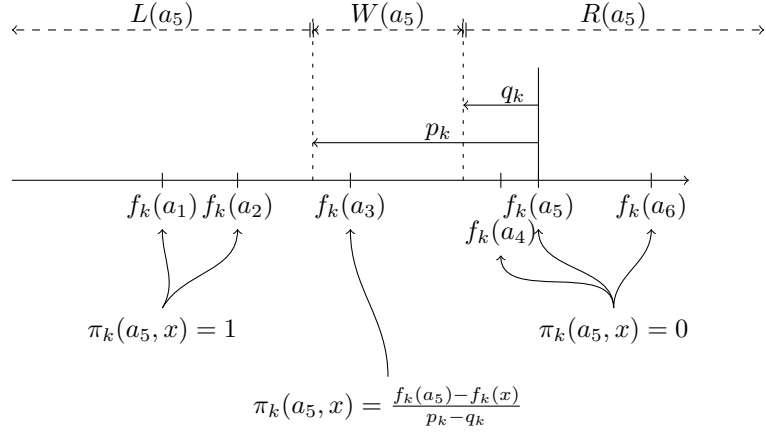
\begin{figure}
		\centering
		\setlength{\unitlength}{1cm}
		\begin{tikzpicture}[domain=0:7,scale=1]
		\draw[->] (-2,0) -- (7,0);
		\draw[->] (5,0.5) -- (2,0.5);
		\draw[->] (5,1) -- (4,1);
		\draw[-] (5,0) -- (5,1.5);
		\draw (0, -0.1) -- (0, 0.1);
		\node (f1) at (0,-0.35) {$f_k(a_1)$};
		\draw (1, -0.1) -- (1, 0.1);
		\node (f2) at (1,-0.35) {$f_k(a_2)$};
		\draw (2.5, -0.1) -- (2.5, 0.1);
		\node (f3) at (2.5,-0.35) {$f_k(a_3)$};
		\draw (4.5, -0.1) -- (4.5, 0.1);
		\node (f4) at (4.5,-0.7) {$f_k(a_4)$};
		\draw (5, -0.1) -- (5, 0.1);
		\node (f5) at (5,-0.35) {$f_k(a_5)$};
		\draw (6.5, -0.1) -- (6.5, 0.1);
		\node (f6) at (6.5,-0.35) {$f_k(a_6)$};
		\put(3.5,0.6){$p_k$};
		\put(4.5,1.1){$q_k$};
		\multiput(2,0)(0, 0.15){15}{\line(0,1){0.05}};
		\multiput(4,0)(0, 0.15){15}{\line(0,1){0.05}};
		\draw[dashed, |->] (4.03, 2) -- (8, 2);
		\node at (6, 2.2) {$R(a_5)$};
		\draw[dashed, |<->|] (4.0, 2) -- (2, 2);
		\node at (3.0, 2.2) {$W(a_5)$};
		\draw[dashed, |->] (1.97, 2) -- (-2, 2);
		\node at (0, 2.2) {$L(a_5)$};
		
		\node (pi_L) at (0, -2) {$\pi_k(a_5, x)=1$};
		\node (pi_W) at (3, -3) {$\pi_k(a_5, x)=\frac{f_k(a_{5})-f_k(x)}{p_k - q_k}$};
		\node (pi_R) at (6, -2) {$\pi_k(a_5, x)=0$};
		\path[->] (pi_L.north) edge [out=60, in=-90] (f1.south);
		\path[->] (pi_L.north) edge [out=60, in=-90] (f2.south);
		\path[->] (pi_W.north) edge [out=90, in=-90] (f3.south);
		\path[->] (pi_R.north) edge [out=90, in=-90] (f4.south);
		\path[->] (pi_R.north) edge [out=90, in=-90] (f5.south);
		\path[->] (pi_R.north) edge [out=90, in=-90] (f6.south);
		\end{tikzpicture}
		\caption[...]{\label{fig:pi_influenced} Illustration of the three different cases for the preference degree $\pi_k(a_5, x)$: $x \in L(a_5)$, $x \in W(a_5)$, and $x \in R(a_5)$}
		
	\end{figure}

One of the observations on which our algorithm for the linear preference function is based, is the following:
There are three different ways in which an alternative $x$ can influence $\phi_k^+(a)$ (see figure \ref{fig:pi_influenced}):
\begin{eqnarray}
	f_k(x)<f_k(a)-p_k&:&\pi_k(a,x) = 1
\\f_k(a)-p_k\leq f_k(x)\leq f_k(a)-q_k&:&\pi_k(a,x)\mbox{~linear~in~}d_k(a,x)
\\f_k(a)-q_k< f_k(x)&:&\pi_k(a,x) = 0
\end{eqnarray}

Hence, except for a window of length $q_k-p_k$, the influence of an alternative $x$ is a default value that depends on the side of the window on which $x$ falls.
We formalize these three regions as follows:
\begin{definition}
		The \emph{window of $a_i$}, denoted $W(a_i)$ is defined as: \[W(a_i)~:=~\{x\in A~|~f_k(x)\in[f_k(a_i)-p_k,f_k(a_i)-q_k]\}\enspace.\]
		We will use $l_i$, respectively $u_i$ to denote the lower and upper bounds of the interval $[f_k(a_i)-p_k,f_k(a_i)-q_k]$.
		We say that an \emph{alternative $x$ is left of $W(a_i)$} if $f_k(x)<l_i$, and that it is right of $W(a_i)$ if
		$f_k(x)>u_i$. The set of all alternatives left, respectively right of $a_i$ will be denoted $L(a_i)$, respectively $R(a_i)$.
\end{definition}

	\begin{example}
		\label{ex:windows}
		We continue Example \ref{ex:prelim}. We illustrate the concept of window by looking at the first criterion ``Sunshine.''
		The alternatives are already ordered according to increasing score for ``Sunshine.'' The window $W(a_1)$ includes all alternatives with a temperature value that is in the
        interval $[5-3,5-1]=[2,4]$. Hence $W(a_1)=\{\}$, $L(a_1)=\{\}$, and $R(a_1)=\{a_1,a_2,a_3,a_4,a_5\}$. The other windows are as follows:
		\[\begin{array}{|c||c|c|c|c|}
		\hline
		a_i     &  [l_i,u_i] & L(a_i)    &   W(a_i)    &   R(a_i)
		\\\hline
		a_1   &   [2,4]   &  \{\}         &  \{\}         & \{a_1,a_2,a_3,a_4,a_5\}
		\\a_2   &   [3,5]   &  \{\}         &  \{a_1\}      & \{a_2,a_3,a_4,a_5\}
		\\a_3   &   [4,6]   &  \{\}         &  \{a_1,a_2\}  & \{a_3,a_4,a_5\}
		\\a_4   &   [4,6]   &  \{\}         &  \{a_1,a_2\}  & \{a_3,a_4,a_5\}
		\\a_5   &   [7,9]   &  \{a_1,a_2\}  & \{a_3,a_4\}   & \{a_5\}
		\\\hline
		\end{array}
		\]
		
	\end{example}
	
	Using these definitions, we obtain the following equality:
	
	\begin{eqnarray}
		\phi_k^+(a)
		&=&\frac{1}{n-1}\left(\sum_{x\in L(a)}1 + \sum_{x\in W(a)} \frac{(f_k(a)-f_k(x))-q_k}{p_k-q_k}
		+\sum_{x\in R(a)}0\right)\nonumber
		\\&=&\frac{1}{n-1}\left(|L(a)|+|W(a)|\times \frac{(f_k(a)-q_k)}{p_k-q_k}-\frac{\sum_{x\in W(a)} f_k(x)}{p_k-q_k}\right)
	\end{eqnarray}

	Hence, in order to compute $\phi^+_k(a)$ for all $a$ efficiently, it is essential to be able to quickly compute $|W(a)|$, $|L(a)|$, and the sum $S(a):=\sum_{x\in W(a)} f_k(x)$ for all alternatives $a$. This can be achieved by a sorting-based method that incrementally computes $|W(a_i)|$, $|L(a_i)|$ and $S(a_i)$, as we show next.

\subsection{Incremental Computation of the Windows $W(a)$}
Let $a_1, a_2, \ldots, a_n$ be the sequence of alternatives, ordered in ascending order of their $k$-th criterion; that is, for all $i=1\ldots n-1$, $f_k(a_i)\leq f_k(a_{i+1})$. After ordering the alternatives, it holds that if $i<j$, $W(a_i)$ will be ``more to the left'' than $W(a_j)$ in the sense that $l_i\leq l_j$ and $u_i\leq u_j$. The following lemma formalizes this observation. We exploit this fact for computing the windows $W(a_i)$ incrementally.
	
	\begin{lemma}
		Let $A=\{a_1,\ldots,a_n\}$ be the set of alternatives ordered in ascending order w.r.t. the $k$-th criterion; i.e., $f_k(a_i)\leq f_k(a_{i+1})$ for all $i=1\ldots n-1$.
		Let $x\in A$, $i<j\in[1,n]$. If $x\in W(a_i)\cap W(a_j)$, then for all $i\leq\ell\leq j$, it holds that $x\in W(a_\ell)$.
		Furthermore, if both $a_r$ and $a_s$ are in $W(a_i)$, then also for all $r\leq\ell\leq s$, $a_\ell\in W(a_i)$.
	\end{lemma}
	
	\begin{example}
		Figure \ref{fig:windows_example} illustrates the observation for the dataset of Example \ref{ex:prelim} and the first criterion.
		
		It is clear that $l_1 \leq l_2 \leq l_3 \leq l_4 \leq l_5$ and $u_1 \leq u_2 \leq u_3 \leq u_4 \leq u_5$, and for instance, because $a_1$ is in $W(a_2)\cap W(a_4)$, $a_1$ is also in $W(a_3)$.

	\begin{figure}[h]
		\centering
		\setlength{\unitlength}{1cm}
		\begin{tikzpicture}[domain=1:12,scale=1]
		\draw[->] (1,0) -- (12,0);
		\draw (5, -0.1) -- (5, 0.1);
		\draw (6, -0.1) -- (6, 0.1);
		\draw (7, -0.1) -- (7, 0.1);
		\draw (7, -0.1) -- (7, 0.1);
		\draw (10, -0.1) -- (10, 0.1);
		\node (f1) at (5,-0.35) {$f_1(a_{1})$};
		\node (f2) at (6, 0.35) {$f_1(a_{2})$};
		\node (f3) at (7,-0.35) {$f_1(a_{3})$};
		\node (f4) at (7,-0.8) {$f_1(a_{4})$};
		\node (f5) at (10,-0.35) {$f_1(a_{5})$};
		\multiput(2,-0.1)(0, -0.15){4}{\line(0,1){0.05}};
		\multiput(4,-0.1)(0, -0.15){4}{\line(0,1){0.05}};
		\draw[dashed, |<->|] (2, -0.6) -- (4, -0.6);
		\node (wa1) at (3, -0.9) {$W(a_1)$};
		\multiput(3,0)(0, 0.15){6}{\line(0,1){0.05}};
		\multiput(5,0)(0, 0.15){6}{\line(0,1){0.05}};
		\draw[dashed, |<->|] (3, 1.0) -- (5, 1.0);
		\node (wa2) at (4, 1.2) {$W(a_{2})$};
		\multiput(4,-0.1)(0, -0.15){6}{\line(0,1){0.05}};
		\multiput(6,-0.1)(0, -0.15){6}{\line(0,1){0.05}};
		\draw[dashed, |<->|] (4, -1) -- (6, -1);
		\node (wa3) at (5, -1.3) {$W(a_3)$};
		\node (wa4) at (5, -1.75) {$W(a_4)$};
		\multiput(7,0)(0, 0.15){6}{\line(0,1){0.05}};
		\multiput(9,0)(0, 0.15){6}{\line(0,1){0.05}};
		\draw[dashed, |<->|] (7, 1.0) -- (9, 1.0);
		\node (wa5) at (8, 1.2) {$W(a_{5})$};

		\end{tikzpicture}
		\caption[...]{\label{fig:windows_example} windows $W_i$ for each alternative $a_i$ on the first criterion from the example}
	\end{figure}
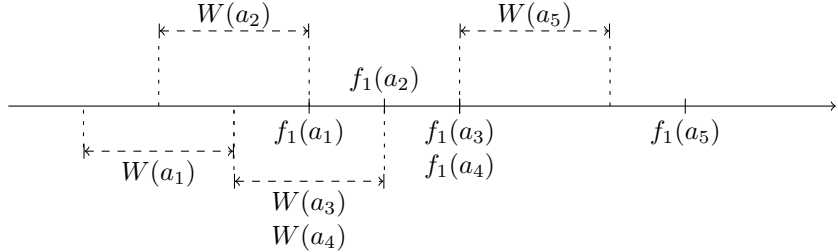
		
	\end{example}

	From the lemma it is clear that if we construct the windows in order from $W(a_1)$ to $W(a_n)$, every time an element $x$ leaves a window, we will never have to reconsider it; that is, if $x\in W(a_i)$, but $x\not\in W(a_{i+1})$, then for all $j=i+1\ldots n$, $x\in L(a_j)$. This gives rise to the following incremental algorithm: when computing $W(a_i)$ from $W(a_{i-1})$, we first shift the leftmost border of the window; we increment the index $\lambda$ until $\lambda=n+1$ or $f_k(a_{\lambda})\geq l_i$, followed by shifting the rightmost border of the window by incrementing index $\upsilon$ until $\upsilon=n$ or $f_k(a_{\upsilon+1})>u_i$. The resulting algorithm is given as Algorithm \ref{algo:window_algo}.

	The key observation in determining the complexity of this procedure is that sorting the dataset on criterion $q_k$ is executed once and takes time $\mathcal{O}(n\log(n))$ using standard out-of-the-box sorting algorithms such as merge-sort.
	Furthermore, lines $2 \rightarrow 9$ are executed only if the test on line $1$ succeeds, which is at most $n+1$ times over \emph{all} incremental computations, because $\lambda$ is incremented by $1$ every time the condition is true.
	In every invocation of Algorithm \ref{algo:window_algo}, the test on line $1$ will fail exactly once.
	Therefore, over \emph{all} iterations the test on line $1$ is executed at most $2n + 1$ times, and lines $2 \rightarrow 9$ are executed at most $n$ times.
	For the second while loop (lines $10 \rightarrow 18$) the same argumentation holds, leading to a time complexity of $\mathcal{O}(n\log(n)+2(n+2n+1))=\mathcal{O}(n\log(n))$.
	
		\begin{algorithm}[H]
 \vspace{0.5cm}	
            \caption{Algorithm to compute $W(a_i)$ from $W(a_{i-1})$ for $i=1...n$}
			\label{algo:window_algo}
			\KwIn{Index $i$,
                \\Indices $\lambda, \upsilon$ such that: if $i=1$, $\lambda = \upsilon = 0$
				\\\quad else, if $i>1$:  $\begin{cases}
					L(a_{i-1}) = &\{a_j~|~j < \lambda\}\\
					W(a_{i-1}) = &\{a_j~|~\lambda \leq j < \upsilon\}\\
					R(a_{i-1}) = &\{a_j~|~\upsilon < j\}
				\end{cases}$\\
				\\and card. $\left|L_{i-1}\right|, \left|W_{i-1}\right|, \left|R_{i-1}\right|$ (if $i=1$ these are resp. 0, 0, and $n$)
			}
			\KwOut{Updated cardinalities $\left|L_{i}\right|, \left|W_{i}\right|, \left|R_{i}\right|$, 
                \\and updated indices $\lambda, \upsilon$ s.t.
				$\begin{cases}
				    L(a_{i}) = &\{a_j~|~j < \lambda\}\\
					W(a_{i}) = &\{a_j~|~\lambda \leq j < \upsilon\}\\
					R(a_{i}) = &\{a_j~|~\upsilon < j\}
				\end{cases}$\\
				}
			
			
			\While{$\lambda \leq n$ and $f_k(a_{\lambda})<l_i$}{
				\eIf{$a_{\lambda}\in W$}{$|W| \leftarrow |W| - 1$ \tcp*{ ${a_{\lambda}}$ moves from $W(a_{i-1})$ to $L(a_i)$ }}{$|R| \leftarrow |R| - 1$  \tcp*{ ${a_{\lambda}}$ moves from $R(a_{i-1})$ to $L(a_i)$ }}
				$|L| \leftarrow |L| + 1$ \tcp*{${a_{\lambda}}$ enters $L(a_i)$}
				$\lambda \leftarrow \lambda+1$ \;
				}
			\While{$\upsilon<n$ and $f_k(a_{\upsilon+1})\leq u_i$}{
				$|R| \leftarrow |R| - 1$ \tcp*{${a_{\upsilon}}$ leaves $R(a_{i-1})$}
				\eIf{$a_{\upsilon}\geq l_i$}{$|W| \leftarrow |W| + 1$ \tcp*{ ${a_{\upsilon}}$ enters $W(a_{i})$ }}{$|L| \leftarrow |L| + 1$  \tcp*{ ${a_{\upsilon}}$ enters $L(a_{i})$ }}
				$\upsilon \leftarrow \upsilon+1$ \;
				}
		\end{algorithm}

\subsection{Incremental computation of $S(a)=\sum_{x\in W(a)} f_k(x)$}

For computing the sum $S(a_i)$ from $S(a_{i-1})$, it suffices to look at the differences between $S(a_i)$ and $S(a_{i-1})$:
\begin{eqnarray}
S(a_i)-S(a_{i-1}) &=& \sum_{x\in W(a_{i})} f_k(x) - \sum_{x\in W(a_{i-1})} f_k(x)
\\&=& \sum_{x\in W(a_{i})\setminus W(a_{i-1})} f_k(x) - \sum_{x\in W(a_{i-1})\setminus W(a_{i})} f_k(x)
\\&=& \sum_{x\in W(a_{i})\cap R(a_{i-1})} f_k(x) - \sum_{x\in W(a_{i-1})\cap L(a_{i})} f_k(x)
\end{eqnarray}
In other words, when incrementally computing the window $W(a_i)$ from $W(a_{i-1})$, we should only keep track of the alternatives leaving $W(a_{i-1})$ (step $3$ of Algorithm \ref{algo:window_algo}), and those entering $W(a_i)$ (step $13$ of Algorithm \ref{algo:window_algo}).

\begin{example}
Consider computing $W(a_5)$ from $W(a_4)=\{a_1,a_2\}$. We start by comparing the leftmost (with respect to the criterion $f_1$ that we are currently evaluating) alternative $a_1$
to $l_5=7$. Since $f_1(a_1)=5<l_5$, we remove $a_1$ from the window and move to the next element. Similarly, $f_1(a_2)=6$ is too small. The first element large enough is $a_3$. Then we start moving the right side of the interval. We know that the right side of the window is at least 1 to the left of the left-side of the interval (in case of an empty interval); hence it is at least $a_2$. Furthermore, we know that the right side of the window should be at least as far to the right as the right side of the window of the previous step. This is again $a_2$. Hence, our first test is if the next element, $a_3$, is also in the window. This is tested by comparing $f_1(a_3)=7$ with $u_5=9$. Since $u_5$ is larger, $a_3$ gets added to the window. Similarly, $a_4$ is added as $f_1(a_4)\leq u_5$. The extension of the window stops as $f_1(a_5)=10$ exceeds the upper bound $u_5$. The new window $W(a_5)$ equals $\{a_3,a_4\}$.

In the course of the computation of $W(a_5)$ from $W(a_4)$, elements $a_1$ and $a_2$ are expelled from the window, while elements $a_3$ and $a_4$ are added. Henceforth, $S(a_4)$ is updated by subtracting $f_1(a_3)$ and $f_1(a_4)$ and adding $f_1(a_3)$ and $f_1(a_4)$.
\end{example}

\subsection{Complete Algorithm}
By combining the incremental computation of the windows and of the sums, we obtain Algorithm \ref{algo:complete_algo} to compute the unicriterion positive flow scores of a set of alternatives.
	
In Algorithm \ref{algo:complete_algo}, we consider the use of queue data structures. These structures act as a \emph{LIFO} (Last In First Out) store. $pop()$ returns and removes the first item of the queue, $append(item)$ adds $item$ to the end of the queue, and $inspect()$ inspects the first element of the queue without removing it. Both $window\_queue$ and $waiting\_queue$ are initialized as queue structures.  $\phi^+$ is initialized as a vector of values to store the computed positive flows.
	
	\begin{algorithm}[H]
        	\caption{Complete algorithm for the computation of unicriterion PROMETHEE II (linear preference function)}
            \label{algo:complete_algo}
\KwData{$q_k$, $p_k$, $f_k$ for all alternatives}
		\KwResult{$\phi^+$ of each alternative}
		initialize($\phi^+$) \tcc*[r]{set up a table to store the $\phi^+$ values}
		W $\leftarrow$ generate\_queue() \;
		R $\leftarrow$ generate\_queue() \;
		sort($f_{k}$) \tcc*[r]{sort the alternatives in ascending order}
		i $\leftarrow$ 2\;
		$\phi^+[1] \leftarrow 0$ \tcc*[r]{first alternative has a pos flow of $0$ }
		
		\While{$i \leq size(f_k)$}{
			$\phi^+[i] \leftarrow \phi^+[i-1]$ \;
			\While{not is\_empty(W) and ($f_k(i)-W.inspect()) \geq p_k$}{
				$window\_item \leftarrow W.pop()$ \;
				$\phi^+[i] \leftarrow \phi^+[i] + \frac{1}{n-1} \frac{p_k -(f_k(i-1) - window\_item)}{p_k - q_k}$ \;
			}
			
			$\phi^+[i] \leftarrow \phi^+[i] + size(W) * \frac{1}{n-1}\frac{f_k(i)-f_k(i-1)}{p_k - q_k}$\;
			
			\While{not is\_empty(R) and ($f_k(i)-R.inspect()) \geq q_k$}{
				$waiting\_item \leftarrow R.pop()$ \;
				$\phi^+[i] \leftarrow \phi^+[i] + \frac{1}{n-1} min(\frac{f_k(i) - waiting\_item - q_k}{p_k - q_k}, 1)$ \;
				\If{$\frac{f_k(i) - waiting\_item}{p_k - q_k} < 1$}{
					$R.append(waiting\_item)$ \;
				}
			}
			$current\_i \leftarrow i$ \;
			\While{$i \leq size(f_k)$ and $f_k(current\_i) = f_k(i)$}{
				$R.append(f_k(i))$ \;
				$\phi^+[i] \leftarrow \phi^+[current\_i]$ \;
				$i \leftarrow i+1$ \;
			}
		}
	\end{algorithm}
	
	\section{Empirical validation}
	\label{sec:results}
	
	We compare the performance of our sorting-based algorithm, that we will call \emph{Sorting Based PROMETHEE} (SBP), with that of the straightforward algorithm based on iterating over all pairs of alternatives\footnote{All code of the algorithms used in the comparisons in this section, as well as scripts for generating the data will be made available online and a link will be included in the final version of the paper.}. Even though from the theoretical analysis it is crystal clear that the $\mathcal{O}(qn\log(n))$ sorting-based method will outperform the $\mathcal{O}(qn^2)$ standard method, we want to illustrate with these experiments how large the actual difference is and to what new problem sizes PROMETHEE II can be scaled. We do not make any comparison of the ranking produced since the sorting-based method is exact and hence the flow scores and ranking produced is always exact. The tests have been executed on a computer with an Intel Core 2 Quad 2.0 GHz and 4 Gb of RAM. The data we use is unicriterion and was generated using a random uniform distribution of values to have as many different values as possible.

\begin{figure}[h]
		\centering
		\includegraphics[width=0.75\textwidth]{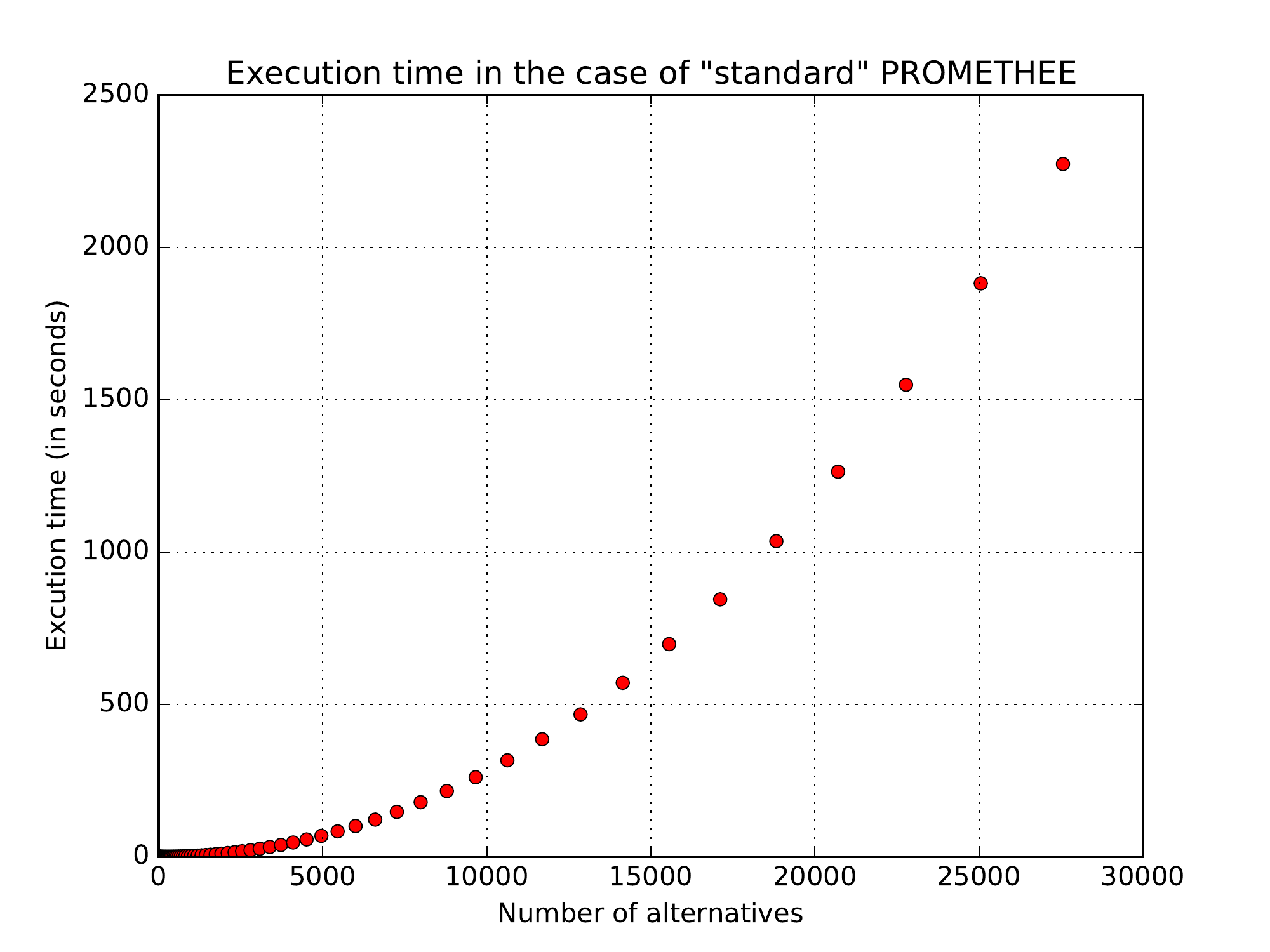}
		\caption{\label{fig:std_promethee}Execution time with respect to the number of alternatives in the case of standard PROMETHEE}
	\end{figure}

\begin{figure}[h]
		\centering
		\includegraphics[width=0.75\textwidth]{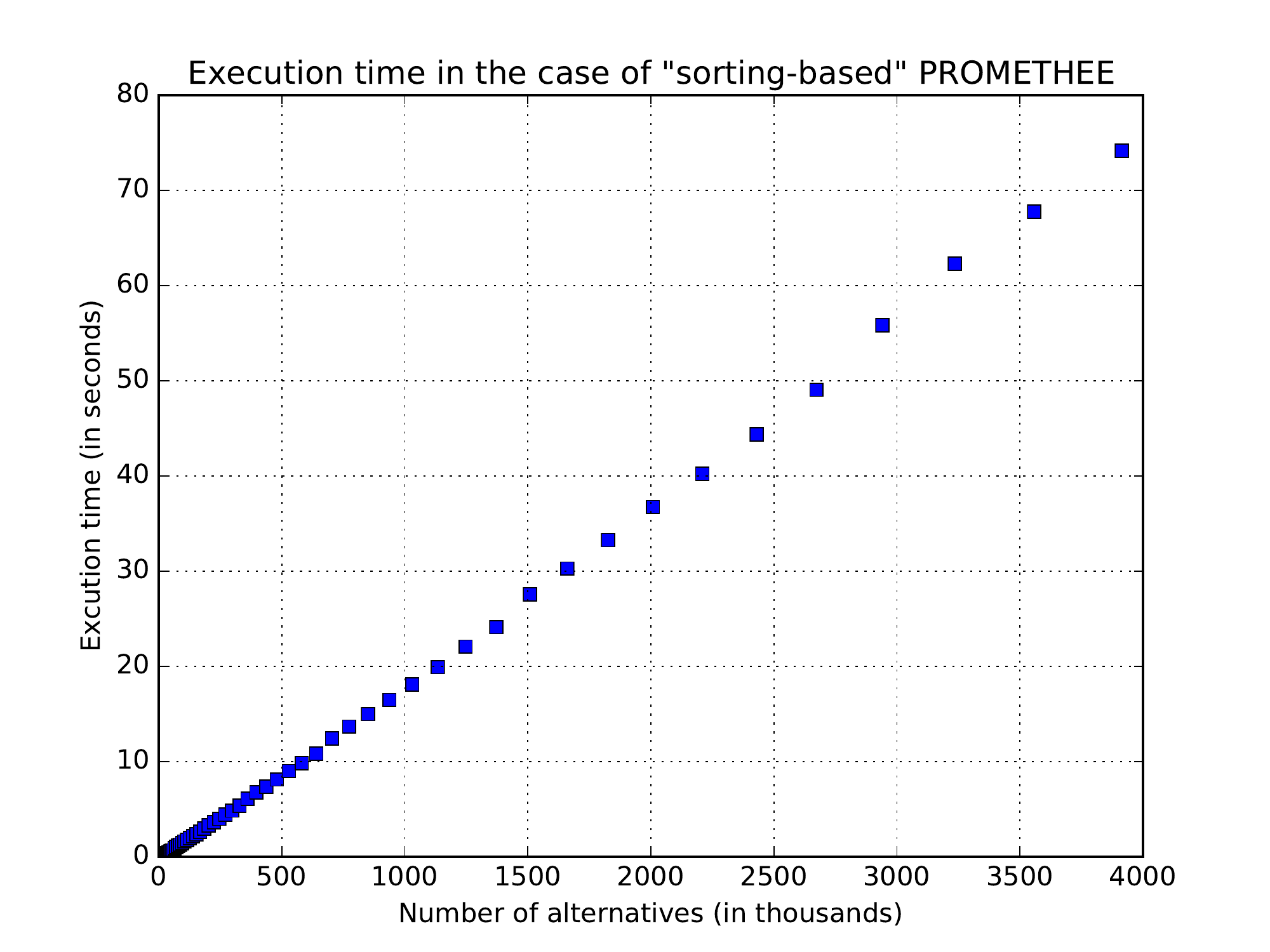}
		\caption{\label{fig:iter_promethee}Execution time with respect to the number of alternatives in the case of sorting-based PROMETHEE}
	\end{figure}

Figure \ref{fig:std_promethee} shows runtime versus number of alternatives for the standard implementation of PROMETHEE, while Figure \ref{fig:iter_promethee} shows the same graph for the sorting-based implementation. The time to compute the unicriterion net flow scores is shown on the Y axis and the number of alternatives to rank on the X axis. Note that the scale on the X axes (number of alternatives) is vastly different because the standard method was not able to scale to larger sizes. To illustrate this difference, the rightmost point in Figure \ref{fig:std_promethee} shows that the standard evaluation method for PROMETHEE for 25000 alternatives takes \emph{over half an hour} while for the sorting-based method this point is on the far left in the graph and takes \emph{less than half a second!}

	Table \ref{tbl:results}, illustrates the computation time's increase in function of the number of alternatives. Every next line the number of alternatives considered is multiplied by $2$. The tables includes the computation times required to compute the unicriterion net flow scores of those sets of alternatives. For standard PROMETHEE, The execution time increases by a factor $4$, as expected due to the complexity in $\mathcal{O}(n^2)$. For the incremental version, however, it increases by a factor only slightly larger than $2$.
	
	For a small number of alternatives, both methods perform satisfactorily, but when the number increases the difference increases rapidly. Indeed, for more than $2^{16}$ alternatives, the standard version needs more than $1$ hour to compute all the scores while the incremental one requires less than a second!
	
	\begin{table}
		\centering
			\begin{tabular}{ | c | c c | }
				\hline
				\# alternatives & Standard PROMETHEE & Incremental PROMETHEE \\
				\hline
				2 & 3e-05 & 5e-05\\
				4 & 7e-05 & 6e-05\\
				8 & 0.00021 & 0.0001\\
				16 & 0.00078 & 0.00017\\
				32 & 0.003 & 0.00032\\
				64 & 0.012 & 0.0006\\
				128 & 0.047 & 0.0012\\
				256 & 0.19 & 0.0024\\
				512 & 0.75 & 0.0048\\
				1024 & 2.98 & 0.0097\\
				2048 & 11.88 & 0.02\\
				4096 & 47.55 & 0.04\\
				8192 & 190 & 0.08\\
				16384 & 775 & 0.17\\
				32768 & 3216 & 0.35\\
				65536 & 12876 & 0.74\\
				131072 &  NA  & 1.56\\
				262144 &  NA  & 3.34\\
				524288 &  NA  & 7.04\\
				1048576 &  NA  & 14.7\\
				2097152 &  NA  & 30.5\\
				4194304 &  NA  & 64.2\\
				\hline
			\end{tabular}
			
		\caption{\label{tbl:results}Execution time in seconds for the standard and incremental PROMETHEE.}
	\end{table}

	\section{Conclusion}
	
	In this paper, we have presented a method that reduces the computation time complexity of PROMETHEE in the case of the linear preference function from $\mathcal{O}(qn^2)$ to $\mathcal{O}(qn\log(n))$. The method is based on an incremental computation after first sorting the alternatives for each criteria. With this new algorithm SBP we have presented, the complexity problem is solved for PROMETHEE; it can now be applied to big sets of alternatives requiring only a small amount of time. The bottom line is: if the dataset can be sorted, PROMETHEE can be applied.
	
	This opens up questions on the interpretation of unicriterion flow scores. Indeed, we have a formulation of $\phi^+(a_i)$ in function of $\phi^+(a_{i-1})$. We can see that with the linear preference function, the difference in scores between $a_i$ and $a_{i-1}$ only depends on the alternatives that were in the windows of $a_i$ and $a_{i-1}$. This could be explored in future research.

Although we have developed our algorithm for the linear preference function, it can easily be extended to the level criterion preference function. The method, however, does not work for the Gaussian preference function. Hence, developing a similarly improved version of PROMETHEE with the Gaussian preference function remains future work as well.

	\bibliography{promethee_n_log_n_tcdva}

\end{document}